\documentclass[%
 aip,
 jmp,%
 amsmath,amssymb,
 reprint,%
]{revtex4-1}

\usepackage{graphicx}
\usepackage{dcolumn}
\usepackage{bm}
\usepackage{xcolor}

\begin{document}


\title{High-stability, high-voltage power supplies for use with multi-reflection time-of-flight mass spectrographs}

\author{P. Schury}
\email{schury@post.kek.jp}
\affiliation{ 
Institute of Particle and Nuclear Studies (IPNS), High Energy Accelerator Research Organization (KEK), Ibaraki 305-0801, Japan
}%

\author{M. Wada}%
\affiliation{ 
Institute of Particle and Nuclear Studies (IPNS), High Energy Accelerator Research Organization (KEK), Ibaraki 305-0801, Japan
}%

\author{H. Wollnik}
\affiliation{%
New Mexico State University, Las Cruces, NM 88001, USA
}%

\author{J-Y. Moon}
\affiliation{%
Institute for Basic Science, Daejeon, Korea
}%

\author{T. Hashimoto}
\affiliation{%
Institute for Basic Science, Daejeon, Korea
}%

\author{M. Rosenbusch}
\affiliation{%
RIKEN Nishina Center for Accelerator-Based Science, Saitama 351-0198, Japan
}%

\date{\today}

\begin{abstract}
Achieving the highest possible mass resolving power in a multi-reflection time-of-flight mass spectrometer requires very high-stability power supplies.  To this end, we have developed a programmable high-voltage power supply that can achieve long-term stability on the order of parts-per-million.  Herein we present the design of the stable high-voltage system and bench-top stability measurements up to 1~kV; the stabilization technique can, in principle, be applied up to 15~kV or more..  We demonstrate that in the $\le$1~Hz band the output stability is on the level of 1~part per million (ppm) during one hour, with only slightly more output variation across 3 days.  We further demonstrate that the output is largely free of noise in the 1~Hz -- 200~Hz band.  We also demonstrate settling to the ppm level within one minute following a 100~V step transition.  Finally, we demonstrate that when these power supplies are used to bias the electrodes of a multi-reflection time-of-flight mass spectrograph the measured time-of-flight is stable on the ppm-level for at least one hour.
\end{abstract}

\maketitle

\bibliographystyle{unsrtnat}

\section{Introduction}

\par A mutli-reflection time-of-flight mass spectrograph (MRTOF-MS) consists of a pair of electrostatic ion mirrors separated by at least one electrostatic lens and a field-free drift region.  The ions oscillate between the mirrors, thereby experiencing a very long flight path.  In principle, each reflection produces a time-focus, where the ion cloud has a maximally short duration, somewhere in the system.  By choosing operational parameters such that after the final reflection the time focus is placed at an ion detector, the MRTOF-MS can achieve extremely high mass resolving powers.    

\par  Although the MRTOF-MS was first proposed in the 1980's \cite{Patent,Wollnik90}, the technology to make the idea useful for reliable, accurate, high-precision mass measurements only became readily available in the past decade.  These devices require efficient ion traps producing maximally low-emittance ion pulses with small energy spread, time-to-digital converters (TDC) with sub-nanosecond accuracy and maximum duration exceeding 10~ms, fast-settling high-voltage switches, and high-stability high-voltage power supplies.  The power supplies remain to be a performance-limiting technical difficulty.

\par As ions spend much of their time near the turning points of the electrostatic mirrors, variations in the potentials applied to the electrodes near the ion turning points can have a very strong affect on an ion's time-of-flight (ToF).  We have observed that a 4~parts-per-million (ppm) change in the voltage of a single electrode can produce a 1~ppm shift in an ion's time-of-flight \cite{Schury2014}.  Such drifts can limit the maximum achievable resolving power by artificially broadening the spectral peak width, while also introducing systematic errors by distorting the peak shape.  Post-processing techniques have been developed to minimize such degradation \cite{Wolf2013,Fischer2018,Schury2018}, but these techniques still require the power supplies to be fairly stable for the duration required to produce a suitable reference measurement.  Achieving the maximum possible performance could require ppm-level stability for durations of $t_\textrm{measure}$$>$10~s.


\par In addition to mass analysis, these devices also are\cite{Wolf2012} and will be\cite{Plass2008,CARIBOU2016,PILGRIM2016} used as online isobar separators for radioactive ion beam facilities.  There are plans to use them for even delivering isomerically pure beams \cite{Dickel2015}.  Voltage stability will present a pressing concern when using these devices as isobar separators to provide high purity beams.  The techniques used to mitigate the affect of ToF drift on spectroscopic analyses are based on post-processing of the spectra and will not be useful to correct drifts in such a high-precision isobar or isomer separator.  While one may envision hourly inspection of the ToF signals and adjustment of the separator, such operation would certainly not be ideal.

\par Presently, some high-voltage power supply companies have begun to bring to market high-stability supplies for use with MRTOF-MS.  However, even the most expensive high-end units offer $\sim$10~ppm long-term stability, temperature coefficients of $\ge$10~ppm/K, and several millivolts of noise in the 0.1~Hz -- 10~Hz range.

\par We have designed and made initial tests of an inexpensive new high-voltage stabilization circuit capable of ppm-level long-term stability, exhibiting low noise and a temperature coefficient $<$5~ppm/K.  They utilize recently available 20-bit DACs and extremely stable voltage references, along with zero-drift, low-noise operational amplifiers (op-amps).  The heart of the design utilizes commercially available photodiodes and high-flux infrared light emitting diodes (LEDs) to emulate transistors with up to 15~kV breakdown voltage.  

\section{Performance criteria}
\par When considering the performance of MRTOF-MS power supplies attention must be given to the short-term ($<$1~s) stability, intermediate-term (1~s -- $\sim$10~minutes) stability, and long-term ($>$10~minutes) stability.  If the power supplies exhibit large fluctuations with frequency components greater than $\sim$0.1~Hz, the aforementioned post-processing techniques will be unable to correct for the induced ToF drift.  To maximize the effectiveness of these post-processing techniques, it is desirable to keep the power supply fluctuations below the level which would result in a statistically significant ToF shift for durations of minutes, to allow ample time to accumulate each reference spectrum (intermediate-term stability).  If the device will be used for isobar (moreso with isomer) separation, it is desirable for the supplies to be sufficiently stable as to preclude the ToF peak moving by more than, perhaps, 50\% of the time-of-flight separator resolution in the course of days (long-term stability).  

\par Since the mass resolving power is only half the time resolving power in an MRTOF-MS, achieving a mass resolving power of $R_\textrm{m}$=10$^5$ requires a time resolving power of $R_\textrm{t}$=2$\times$10$^5$. The most sensitive electrode in our full-scale system exhibits $\Delta$$t$=0.26~ppm time drift from a 1~ppm voltage drift \cite{Schury2014}. Assuming all other voltages were perfectly stabilized and further assuming a voltage distribution in which the MRTOF had an infinite inherent resolving power, the voltage applied to the most sensitive electrode could not exceed a voltage variance $\sigma$($V$)/$V$$<$19~ppm in order to achieve $R_\textrm{m}$=10$^5$.  Achieving $R_\textrm{m}$=5$\times$10$^5$ would require $\sigma$($V$)/$V$$<$3.8~ppm.  As the fluctuations in the bias applied to each electrode independently contribute to the width, while the inherent resolving power is not infinite, the stability requirements become even more stringent.

\par One strategy for limiting voltage fluctuations is to use a resistor-capacitor (RC) low-pass filter.  This has some limitations, however.  Among capacitors rated for 5~kV the maximum capacitance is 10~nF, unless resorting to large, oil-filled capacitors.  The resistor must be chosen so as to not create large Johnson noise ($\approx$4~$\mu$V/$\sqrt{\textrm{Hz}}$/M$\Omega$ at room temperature).  In the case of a 5~kV potential, such a criterion would permit the use a 1~G$\Omega$//10~nF low-pass filter to achieve a time constant $\tau$=10~s while keeping the contribution of Johnson noise below 1~ppm.  Unfortunately, the leakage current of high-voltage capacitors can readily fluctuate on the level of nanoamps \cite{HorowitzHill}.  A fluctuation of $\Delta$$I_{leak}$=10~fA would already cause a 1~mV variation in voltage drop across a 1~G$\Omega$ resistor -- corresponding to 1~ppm at 1~kV.

\par On the other hand, a 1~M$\Omega$//10~nF low-pass filter would not be so affected by these deleterious effects, but has only a $\tau$=10~ms time constant.  Of course, larger capacitors could be implemented, however achieving $\tau$=10~s or more can only be accomplished with large, oil-filled capacitors, which are available with capacitances exceeding 10~$\mu$F but also pose a serious safety hazard.  

\par In addition to the dangers posed by high-capacitance high-voltage capacitors, very long time constant low-pass filters have other downsides, too.  During the initial tuning of an MRTOF-MS, the potentials applied to each electrode should be scanned within a few volt range in order to find the optimal parameters.  Stabilization based on long time constant RC low-pass filtering can require a pause of as much as 10~minutes between each step in such a scan, to allow the voltage to sufficiently stabilize, making initial tuning of the MRTOF-MS a laborious task. 

\par In one effort to overcome the limitations of presently available power supplies, the use of a long time-constant RC filter in combination with a software proportional-integral regulation circuit has been suggested \cite{Wienholtz2019}.  In that case, it was demonstrated that a large RC filter does significantly mitigate fluctuations.  However no quantitative results of the improved settling time were given.  In contrast to this, we have utilized an active negative feedback ``high-voltage op-amp'' to achieve the improved power supply performance needed for MRTOF-MS.

\begin{figure}[ht]
	\centering
	 \includegraphics[width = 3.35in]{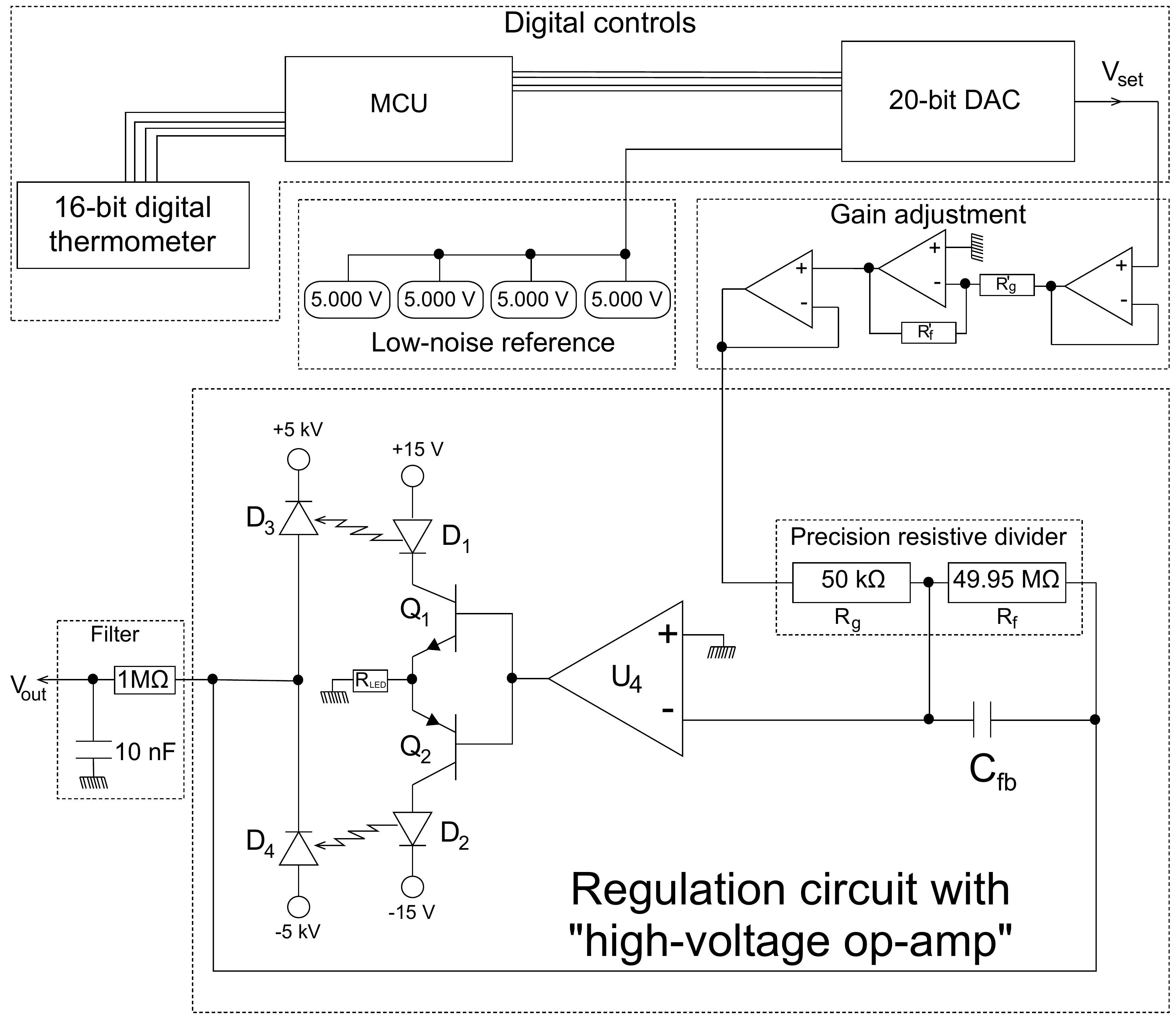} 
	\caption{\label{figConcept}  Design of the high-voltage stabilization circuit.  The regulation block makes use of a "high-voltage op-amp" built from photodiodes (D3 and D4) and infrared LEDs (D1 and D2).  Complementary transistors Q1 and Q2 balance the intensity of infrared light emitted from D1 and D2 based on the output of error amplifier U4.  A precision resistive divider offers negative feedback with a gain of $g$=$-999$.  A micro control unit (MCU) is used to monitor the temperature using a 16-bit digital thermometer and to program the 20-bit DAC.  The DAC reference voltage uses four precision, low-noise voltage references in parallel to halve the inherent noise and average out the thermal and long-term drifts of the references.  The gain adjustment circuit allows for the full-range output to be modified while using a fixed gain for the ``high-voltage op-amp'' circuit, and also provides an opportunity to balance the temperature coefficient of the precision resistive divider.
	} 
	\vspace{-1.5mm}
\end{figure}

\section{High-voltage stabilizer design}

\par In order to achieve a fast-settling, high-precision, long-term stable high-voltage power supply for driving our MRTOF-MS, we have extended the concept laid out in the Voltage Multipliers, Inc, application note ``AN-0300 - High Voltage Op-Amp Application Using Opto-Couplers" \cite{VMI_AN}, which describes the use of infrared LEDs and high-voltage photodiodes to emulate an operational amplifier with up to 15~kV power supply span.  A conceptual circuit diagram of our design, separated into functional blocks, is given in Fig.~\ref{figConcept}.




\par A microcontroller (MCU) is used to read the system temperature as measured by a 16-bit thermometer (Analog Devices ADT7310) and to program a 20-bit DAC (Analog Devices AD5791).  The DAC reference voltage, V$_\textrm{ref}$=5~V, is produced using four parallel units of LTC6655-5 to halve the inherent noise level.  The DAC output, $V_\textrm{set}$, can span a range of -V$_\textrm{ref}$ to +V$_\textrm{ref}$.  A gain adjustment block rescales this range by a factor of -R$^\prime_\textrm{f}$/R$^\prime_\textrm{g}$.  This adjusted DAC output is the set voltage for the ``high-voltage op-amp" block.

The high-voltage op-amp block is derived from the aforementioned application note (AN-300).  In this block, a pair of complementary phototransistors in a push-pull configuration is constructed from basic components.  With reference to Fig.~\ref{figConcept}, an npn phototransistor is built from D$_1$, D$_3$, and Q$_1$ while a pnp phototransistor is built from D$_2$, D$_4$, and Q$_2$.  In the system presented herein, D$_1$ and D$_2$ are high-intensity infrared LEDs (Osram SFH4235), D$_3$ and D$_4$ are high-voltage photodiodes (Voltage Multipliers Inc.~OZ100SG), and Q$_1$ and Q$_2$ are complementary bipolar junction transistors (2N4923 and 2N4920, respectively).  The presently used OZ100SG photodiodes have a reverse breakdown voltage of 10~kV and a maximum reverse current of $\sim$1~mA; if higher voltages were needed the OZ150SG is a drop-in replacement that could be used to extend the supply span to 15~kV.  In principle, stacking multiple photodiodes could extend the voltage span further.

In a manner akin to that often used to linearize complementary symmetrical push-pull output stages, the output is fed back to U$_4$ (Analog Devices AD8676) in a negative feedback loop.  A precision resistive divider (Caddock HVD5 series, R$_\textrm{g}$=50~k$\Omega$ and R$_\textrm{f}$=49.95~M$\Omega$) connects the high-voltage output and adjusted DAC output to the inverting input of U$_4$, which then plays the role of error amplifier producing a gain of $g$=$-999$ for the high-voltage op-amp block.


To prevent oscillations, a 1~nF ceramic disc capacitor (C$_\textrm{fb}$ in Fig.~\ref{figConcept}) rated for 10~kV$_\textrm{DC}$ (Murata DHRB34A102M2BB) is placed in parallel to R$_\textrm{f}$, making an $f_\textrm{c}$=3~Hz low-pass filter.

\begin{figure}[th]
	\vspace{-6mm}
	\centering
	 \includegraphics[width = 3.45in]{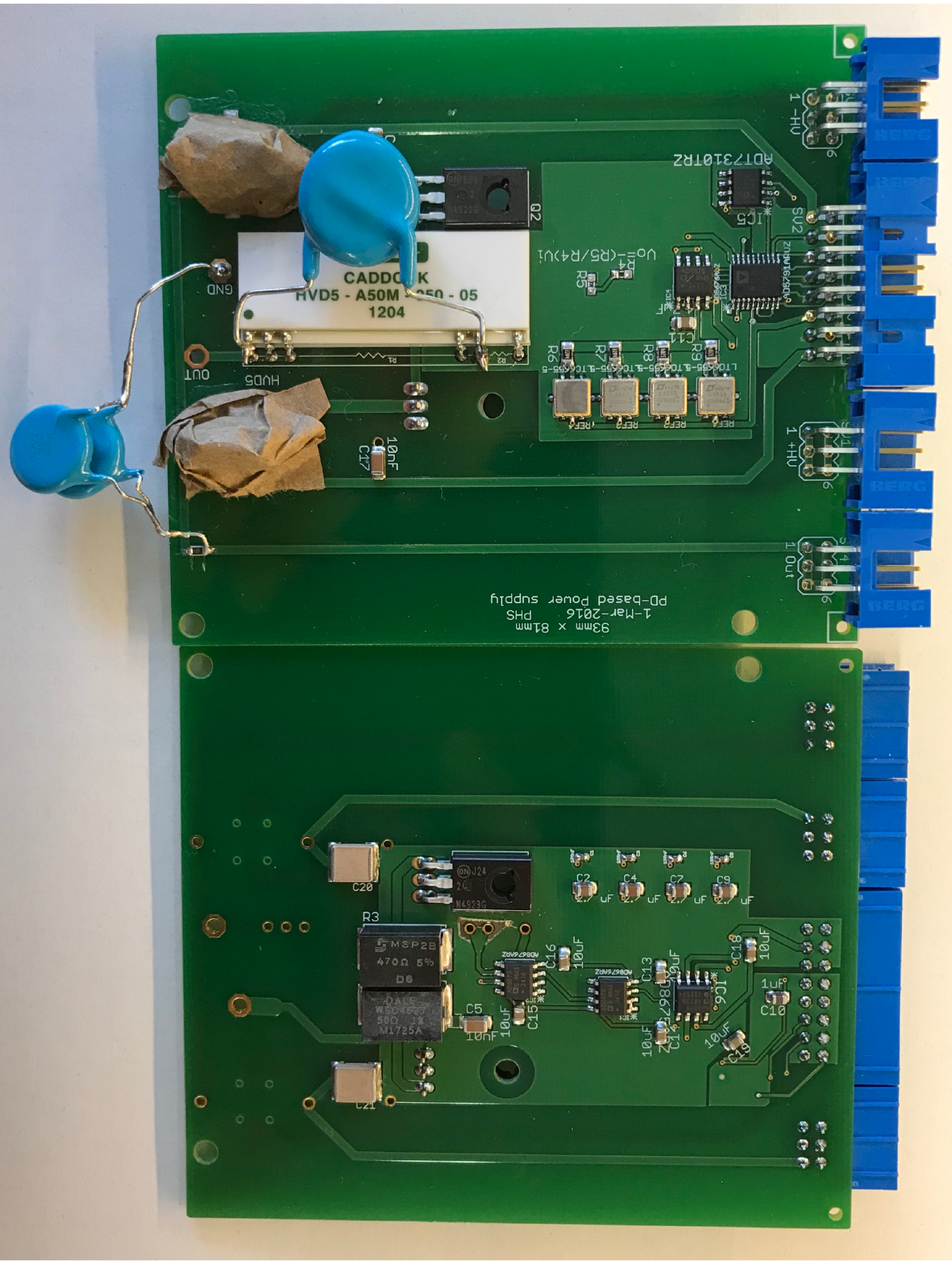} 
	\caption{\label{figPhoto} Photograph of the prototype printed circuit board used in this work.  The photodiodes, covered by brown paper to shade them from ambient light sources, are mounted atop high-flux infrared LEDs.     
	} 
\end{figure}

\par  The inclusion of the gain adjustment block may seem superfluous at first blush, but it actually serves some useful purposes.  The precision resistive divider has a ratio temperature coefficient of $\pm$5~ppm/K.  By careful selection of the resistors R$^\prime_\textrm{f}$ and R$^\prime_\textrm{g}$, based on their measured temperature coefficients, it is possible to balance the temperature coefficient of the precision voltage divider and produce a net gain with almost zero temperature coefficient.  Additionally, the optimal potential which should be applied to many of the MRTOF-MS electrodes is much smaller than 5~kV.  At a gain of $g$=-999 the minimum output step size is 9.5~mV, which may be inconveniently large in cases where the optimal bias potential is $e$.$g$ $\approx$500~V.  The use of the Gain adjustment block in Fig.~\ref{figConcept} allows the full-scale range to be selected so as to be optimal for each electrode.  Bipolar operation allows the MRTOF to be used with positive or negative ions.



\par To remove any fluctuations at frequencies far exceeding our ability to precisely measure, a 1~M$\Omega$//10~nF low-pass filter has been added to the output.  While the large output impedance may introduce some extra Johnson noise, and fluctuation in the capacitor's leakage current may produce some voltage fluctuations, this was determined to be an acceptable compromise.  Additionally, the inclusion of such a low-pass filter simplifies the optional addition of large, oil-filled high-voltage capacitors for long time-constant low-pass filters in the future, if yet higher intermediate-term stability becomes necessary.

\par A 16-bit digital thermometer (Analog Devices ADT7310) was installed near the DAC to monitor the local temperature.  Monitoring the temperature allows for indication of failures, such as \emph{e.g}.\ oscillations that can cause the LED driving system to dramatically heat up.  It also is foreseen that such thermal monitors could be useful in a future active thermal stabilization of the power supplies. 

\section{Testing procedures and results}
\par A number of benchtop tests were performed to characterize the performance of the circuitry shown in Fig.~\ref{figPhoto}.  These tests included settling time, reproducibility, linearity, an analysis of thermal effects, measurements of long-term, intermediate-term, and and shot-term stability, and determination of noise in the 1~Hz -- 200~Hz range.  For these measurements, a minimal setup was utilized.  An Agilent 3458A 8.5-digit digital multimeter was used to digitize the output voltage.  A GPIB-to-Ethernet interface (Prologix, LLC) allowed interfacing with the 3458A using TCP/IP.  Rather than $\pm$5~kV, the supply rails were initially set to +1~kV and 0~V to protect the 3458A from possible damage due to overvoltage.

\par A Raspberry Pi \cite{Raspi} Model 3 computer was implemented as the micro controller unit (MCU) of Fig.~\ref{figConcept} while also serving as a rudimentary data acquisition and control system.  The Raspberry Pi's GPIO interface was used to communicate with the power supply, both setting the DAC output and reading from the thermometer.  Similarly, a computer code running on the Raspberry Pi was used to set the digitization rate of the 3458A and record the digitized data, interfacing via TCP/IP.

\par However, before performing any of these studies, to properly account for the device's 1~M$\Omega$ output impedance, the multimeter's precise input impedance had to be determined.  The voltage before the filter was made to be 100.000~V (as measured by the 3458A).  Finally, with the true output being 100.000~V, the voltage after the filter, measured using the 3458A, was found to be 90.869~V.  Hereon, all stated voltages have been adjusted from the measured value to account for this.

\subsection{Settling time}
\par
\begin{figure}[bth]
	\centering
	 \includegraphics[width = 3.35in]{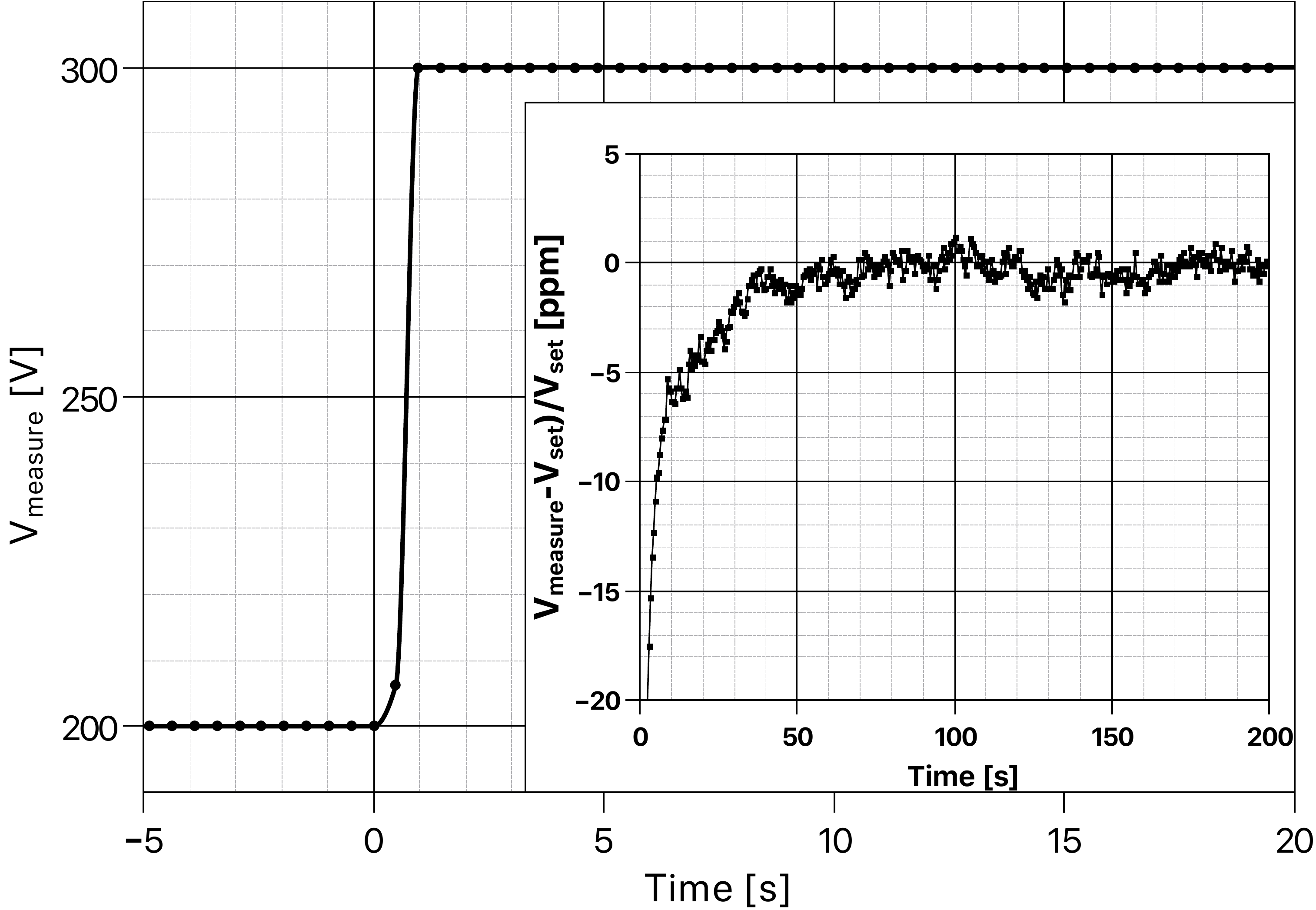} 
	\caption{\label{figSettling} The power supply settles on the ppm level within 50~s after a 100~V change in the output.  The digitization rate was 1.7~Hz.    \vspace{-5 mm}
\vspace{-3 mm}
	} 
\end{figure}

 In order to tune the MRTOF-MS in a reasonably finite duration, as previously noted, it is valuable for the power supply to settle reasonably quickly following a step change in the output.  To test the settling time, the output was first set to 200~V and left to equilibrate for several minutes.  The data acquisition was started with a digitization rate of 1.7~Hz.  After approximately one minute of data acquisition, while continuing the signal digitization, the DAC output was changed to produce a 300~V output from the power supply.  The result, shown in Fig.~\ref{figSettling} zoomed to show detail of the final 20~ppm of the transition, is that the system settles on the ppm-level within approximately 1~minute.  Within 10~s, the system having settled to within $\approx$5~ppm of the new set value, it would be reasonable to begin production of a ToF spectrum for MRTOF-MS tuning.

\subsection{Reproducibility and linearity}
\par Two further characterizations of a precision power supply are the linearity and reproducibility.  These qualities determine the accuracy which can be anticipated when \emph{e.g}. scaling the voltages in the MRTOF-MS.  With a high degree of accuracy, the potentials should be able to be scaled to change the ion energy without need for a time-consuming retuning of the voltages.  Additionally, it is useful to understand the linearity.

\par To test the reproducibility of the system, the DAC output was switched between values appropriate to produce 200~V and 300~V.  In light of the measured settling time, the power supply was given 60~s to stabilize after each transition, after which 20 consecutive digitizations of the output were made at 1.7~Hz and the average and standard deviation recorded.

\begin{figure}[th]
	\centering
	 \includegraphics[width = 3.4in]{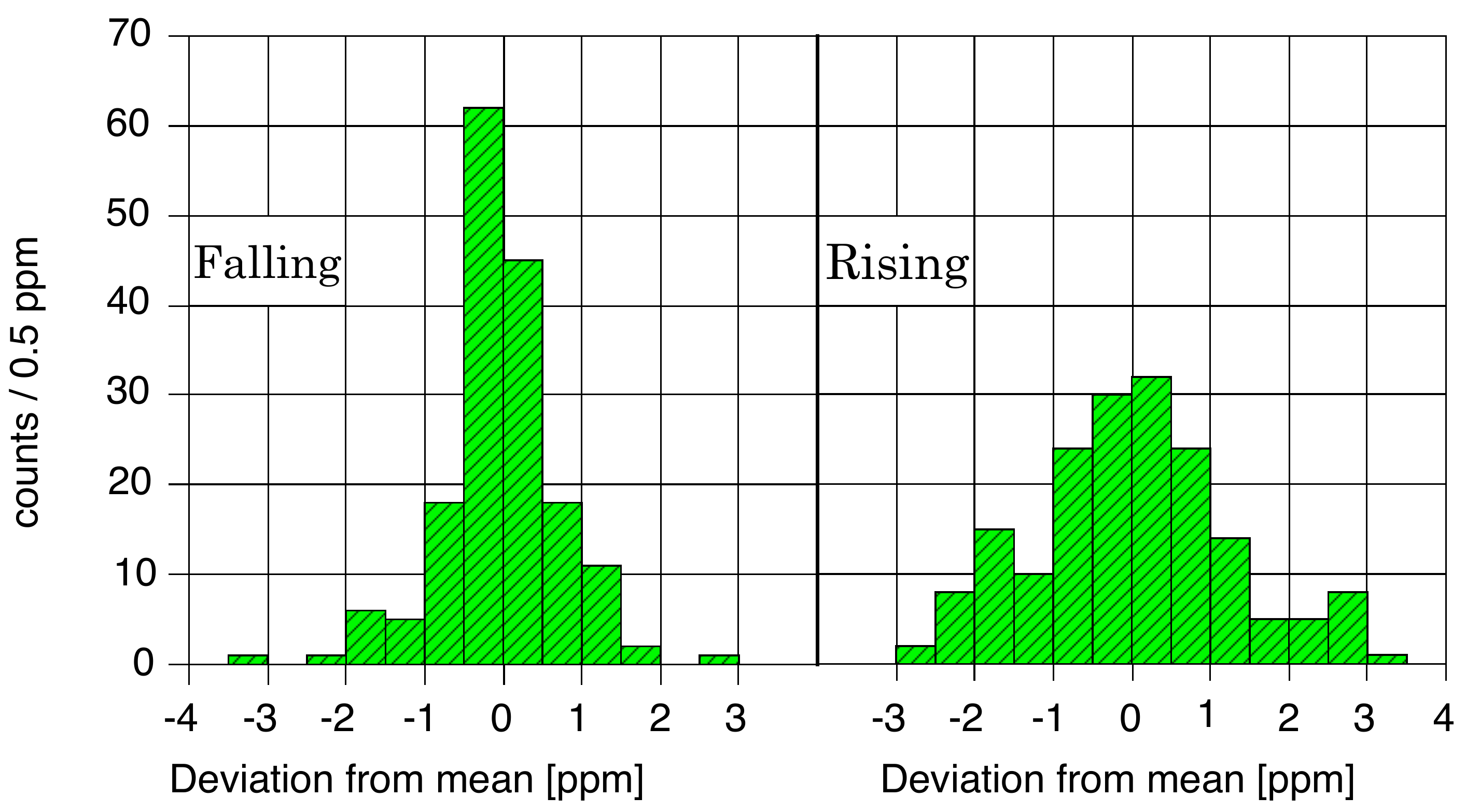} 
	\caption{\label{figRep} Results of reproducibility test. The output was switched back and forth between 200~V and 300~V.  Histograms of the deviation from the desired voltage after 60~s of settling time show the desired output can be reliably obtained within a few parts per million for both rising and falling transitions.    
	} 
\end{figure}

\par Over an $\approx$8~hour period, 180~transition cycles were made.  Histograms showing the measured deviation from desired outputs are shown in Fig.~\ref{figRep}.  For reasons unknown the reproducibility appears to be better following a falling transition (``V$_\textrm{out}$=200~V").  The reproducibility in either case is on the ppm-level, indicating that it should be able to reliably reproduce the desired voltages with the level of accuracy that would be desired for MRTOF-MS systems of the highest conceivable resolving powers.


\par The linearity was tested in DAC code increments of 200.  After sending a new DAC code, the system was allowed 10~s to stabilize, after which 10 consecutive digitizations of the output were made at 1.7~Hz and the average and standard deviation recorded.  Upon reaching the maximum DAC code, the measurement was repeated in reverse.
\par
\begin{figure}[th]
	\centering
	 \includegraphics[width = 3.4in]{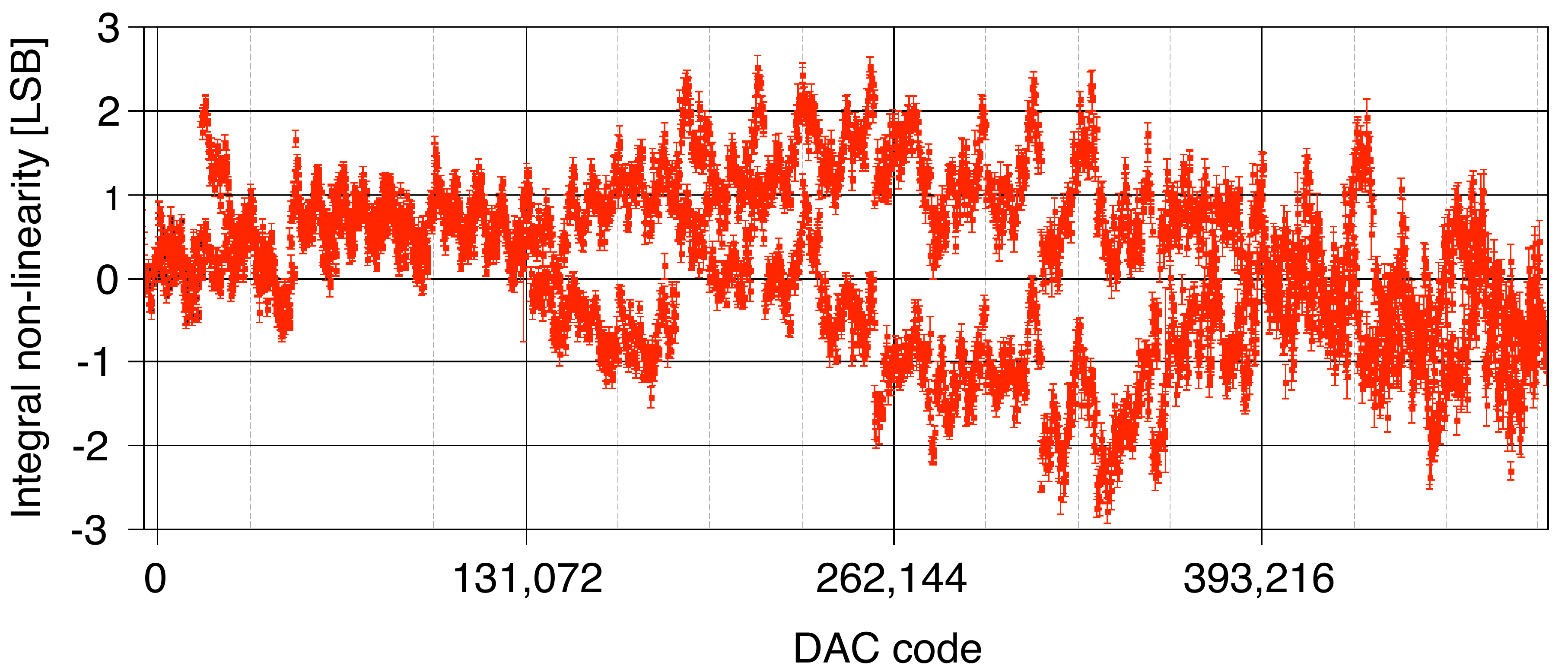} 
	\caption{\label{figLin} Results of linearity test.  The integral non-linearity is compatible with the linearity of the AD5791 20-bit DAC at the heart of the system.  See Fig.~7 of the Analog Devices AD5791 data sheet (Rev. D) for comparison.  In this measurement the LSB voltage was 715~$\mu$V.     
	} 
\end{figure}

 Figure~\ref{figLin} shows the result of the linearity measurement.  The integral non-linearity was determined by dividing the residual of the best-fit line of $V_\textrm{read}$ as a function of $V_\textrm{set}$ by the least significant bit (LSB) voltage.  In this measurement the LSB voltage was 715~$\mu$V.

\subsection{Thermal effects}
\par There are several thermal effects which could degrade the performance of the device.  The Seebeck effect, whereby voltages are induced by thermal gradients at an interface of dissimilar materials ($\emph{e.g}$. solder joints), as previously mentioned, is rather difficult to evaluate in such a device where multiple components may exhibit self-heating.  However, the effect should be below the 100~$\mu$V level based on the components used, and should not produce dynamic changes in the output once the device has achieved thermal equilibrium.  It is possible that these effects will noticeably inhibit reproducibility between extended powered off periods. 

\begin{figure}[b]
	\centering
	 \includegraphics[width = 3.45in]{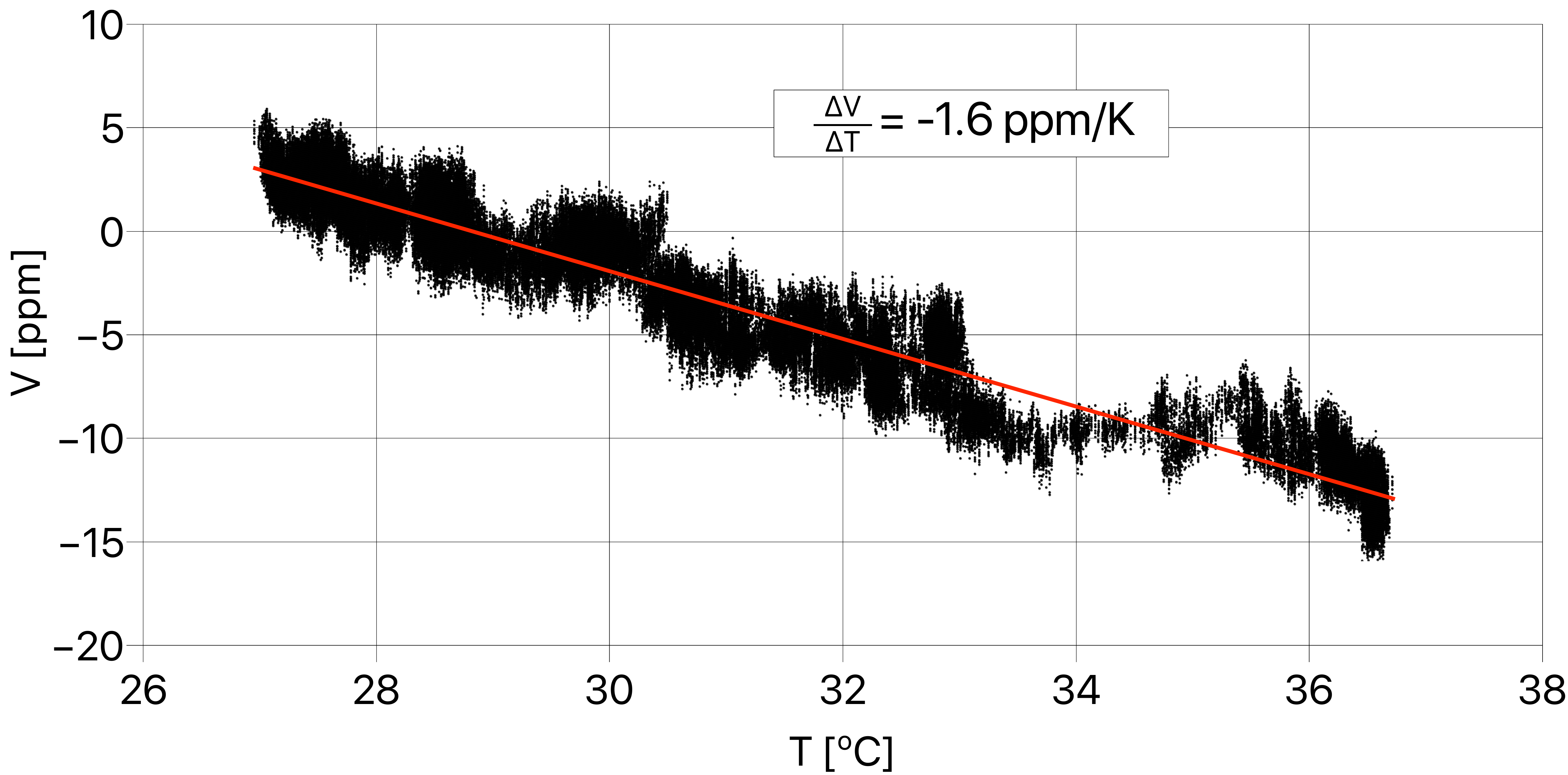} 
	\caption{\label{figThermalDVDT} Result of measurement of the temperature response of the device, showing a temperature coefficient of -1.6 ppm/K.
	} 
\end{figure}

\par The temperature coefficients of the voltage reference and of the gain-setting resistors will combine to produce a gain temperature coefficient affecting the output, which will produce a similar relative shift at all output voltages.  The data sheet for the LTC6655-5 voltage reference lists the typical temperature coefficient to be 1~ppm/K, while the Caddock HVD5-series 1000:1 voltage divider has a stated ratiometric temperature coefficient of $\pm$5~ppm/K.  A passive measurement of the temperature coefficient was performed by placing the device in an environ without temperature stabilization and setting the output to 300~V.  The temperature and voltage were then measured over the course of 2 days.  From the result, shown in Fig.~\ref{figThermalDVDT}, a temperature coefficient of $\Delta$$V/\Delta$$T=-1.6$~ppm/K could be ascertained.  However, the 1~M$\Omega$ resistor (Vishay P1206Y1004BNTA) in the RC filter block has a temperature coefficient of $\pm$10~ppm/K.  As the input impedance of the 3458A is 10~M$\Omega$ with near zero temperature coefficient, the measured temperature coefficient therefore has a systematic uncertainty of $\pm$1~ppm/K due to the filter block.

\par However, in addition to the multiplicative effects from such temperature coefficients, a temperature-dependent voltage offset is also possible.  The most likely source of such an offset in this device would be the AD8676 op-amp (U$_4$ in Fig.~\ref{figConcept}) used to drive the $g=-999$ photodiode-based amplifier.

\par
\begin{figure}[t]
	\centering
	 \includegraphics[width = 3.55in]{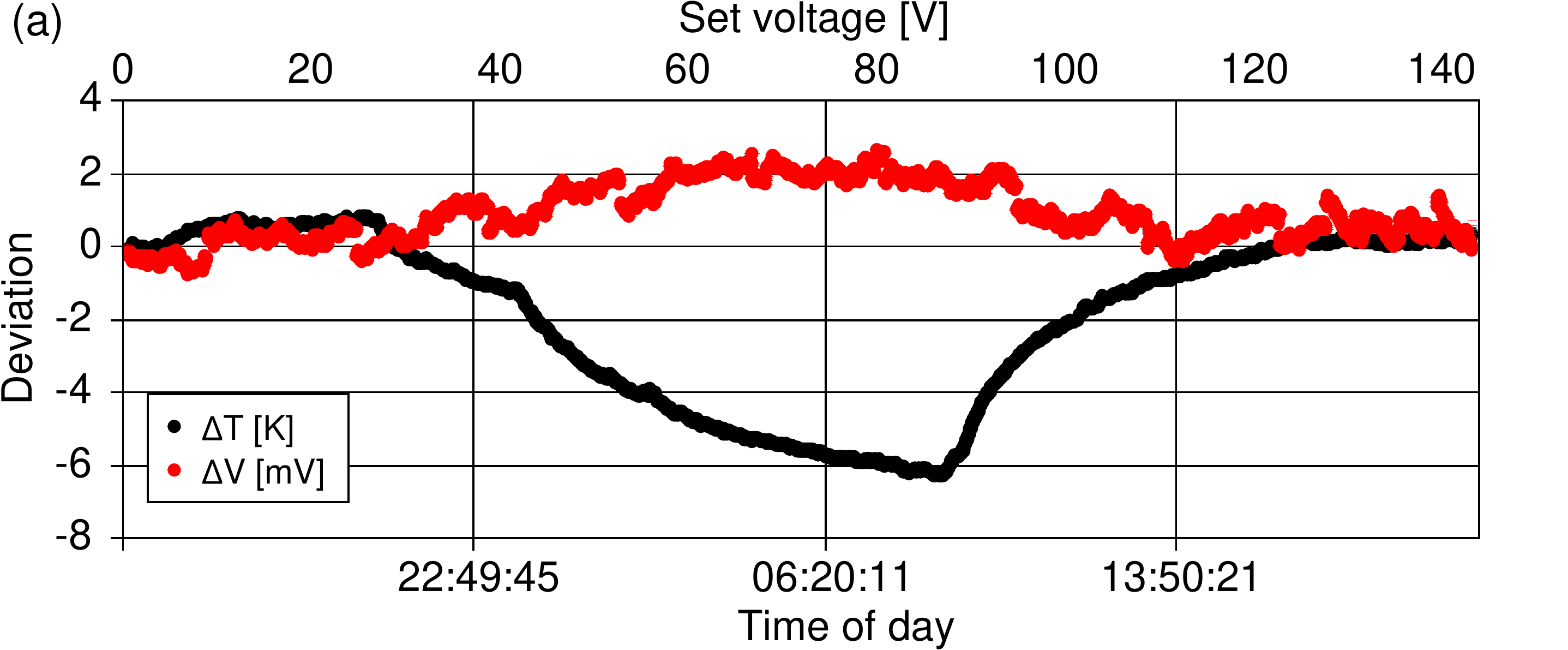} 
\newline

	 \includegraphics[width = 3.35in]{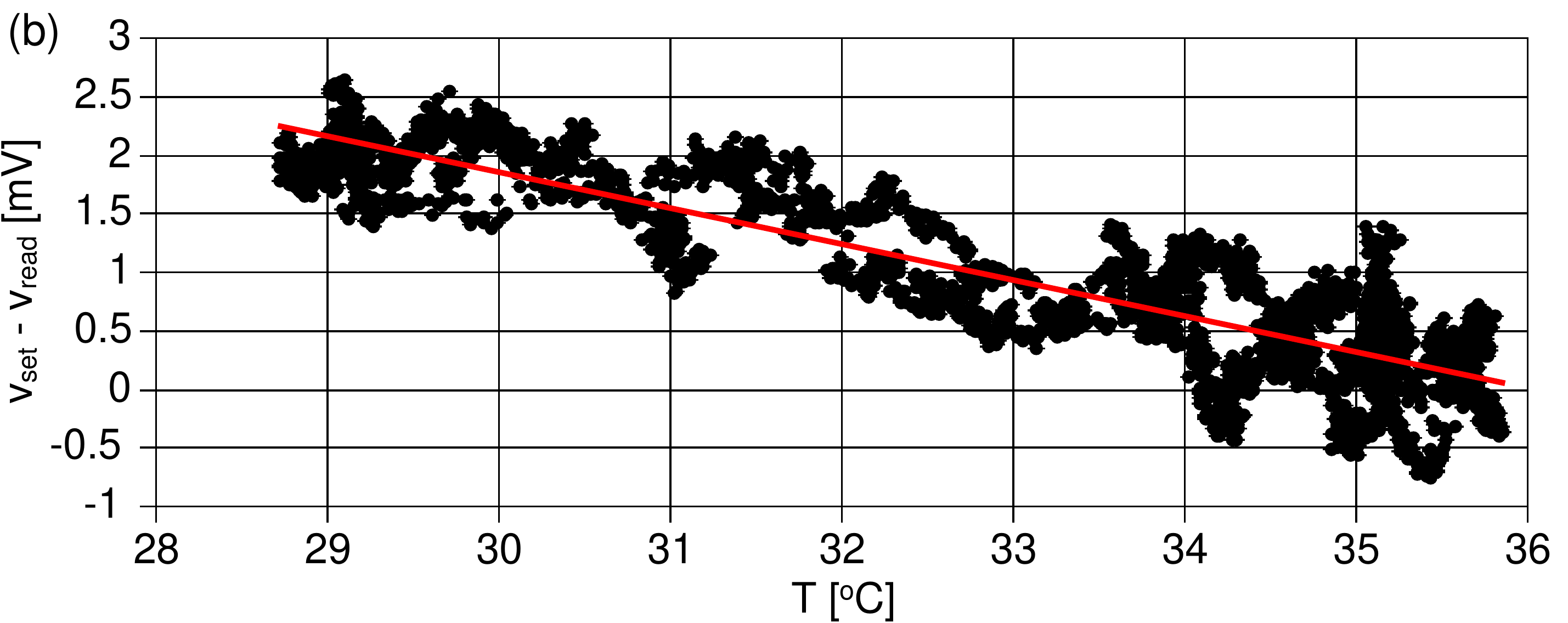} 
	\caption{\label{figThermal} (a) Result of a slow scan of the DAC output during which the temperature of the room changed.  (b) A plot of the difference between set and read voltages.  The red line is the line of best fit, which indicates the existence of a -307(2)~$\mu$V/K temperature dependent offset voltage.  This is consistent with the listed offset voltage of the AD8676 (typically 0.2~$\mu$V/K, maximum 0.6~$\mu$V/K) used to drive the $g$=$-$999 photodiode-based amplifier.  
	} 
\end{figure}

 In order to separate the temperature coefficient of the voltage offset from the gain temperature coefficient, the DAC output was slowly scanned -- similar to the linearity test, but in smaller steps -- over a one day period without thermal stabilization.  The thermometer reading decreased by 6~K and then recovered again during the measurement.  The resultant data is plotted in Fig.~\ref{figThermal}.

\par  In Fig.~\ref{figThermal}a it can be seen that the maximum deviation from initial temperature is well-aligned with the maximum absolute deviation between $V_\textrm{set}$ and $V_\textrm{read}$, near $V_\textrm{set}$=80~V.  
By plotting the absolute voltage deviation as a function of the temperature, as shown in Fig.~\ref{figThermal}b, a linear correlation is seen, with a slope of $-307(2)=\mu$V/K.  This is the voltage offset temperature coefficient.

\par At V$_\textrm{set}$=300~V the temperature offset coefficient would be responsible for $\Delta$$V/\Delta$$T$=$-1.0$~ppm/K.  As such, we can infer that the true gain temperature coefficient is a mere $\Delta$$V/\Delta$$T=-0.6(1.0)$~ppm/K.

\subsection{Long-term stability}
\par The primary motivation for development of the new power supply being to provide extremely stable voltages for MRTOF-MS, the intermediate and long term stability were studied in-depth.  To exclude thermal effects, the device was placed in a temperature controlled room.  The DAC was programmed so as to produce a 400~V output.  The Agilent 3458A was set to average for 30 power cycles (600~ms at 50~Hz mains power) resulting in a digitization rate of $\approx$1.7~Hz.  After allowing the device a few hours to achieve thermal equilibrium, the digitized output was recorded along with the temperature of the digital thermometer over a period of multiple days.

\par
\begin{figure}[t]
	\centering
	 \includegraphics[width = 3.45in]{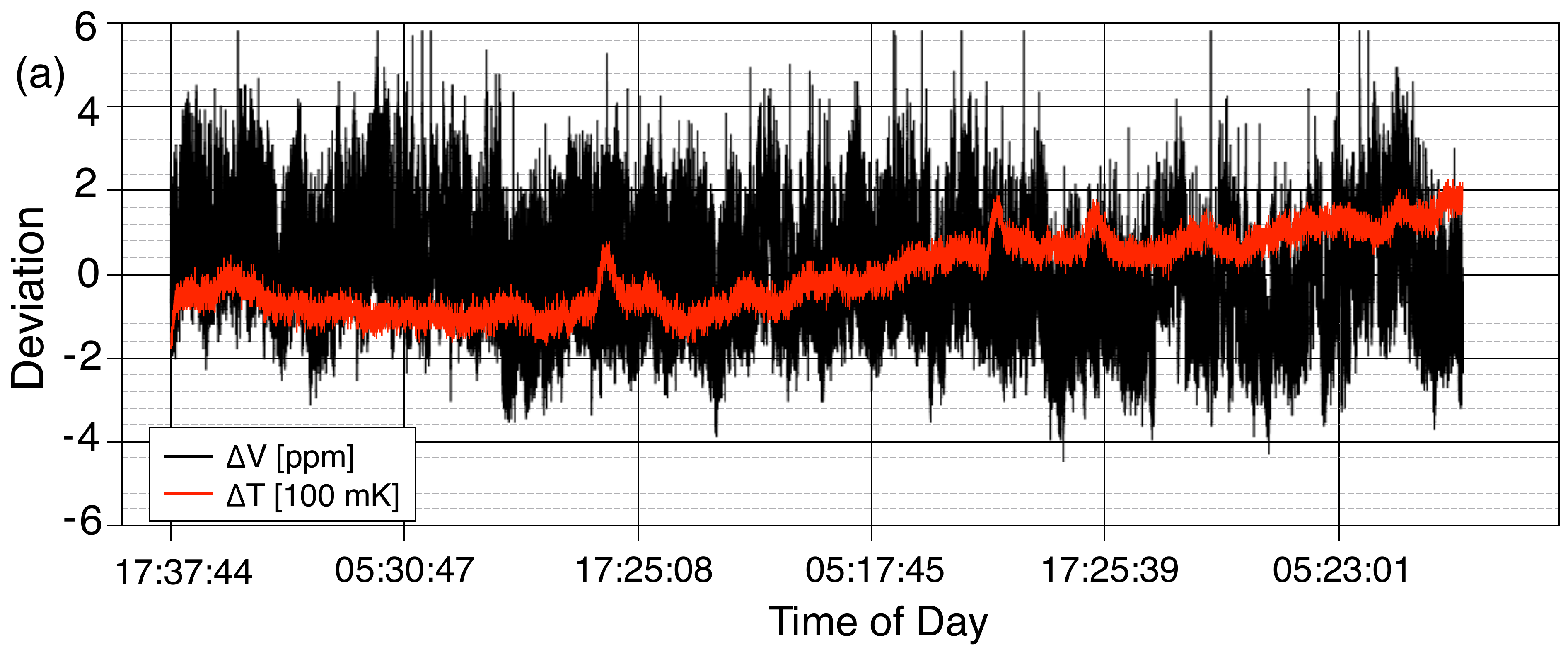} 
	 \newline
	 
	 \includegraphics[width = 3.5in]{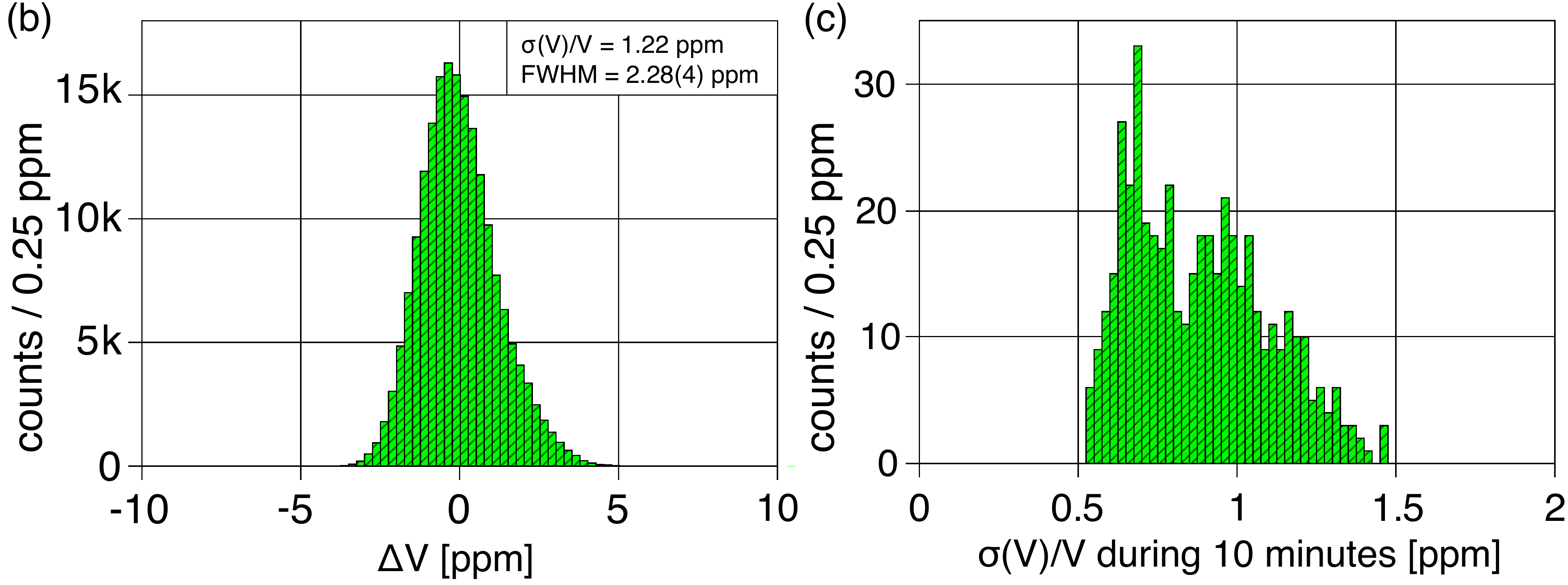} 
	\caption{\label{fig60Hours} (a) Stability of one voltage stabilizer during 3~days of operation at V$_\textrm{out}$=400~V.  The voltage digitization rate was 1.7~Hz, while the temperature was recorded every 5~s.  (b) Histogram showing the voltage stability over the course the measurement.  A Gaussian fit (red curve) shows a full-width at half-maximum stability of 2.28(4)~ppm. (c) Histogram of standard deviation across all consecutive 10~minute intervals within the data shown in the top panel.  The mean standard deviation was 0.88(21)~ppm. 
	} 
\end{figure}

 As shown in Fig.~\ref{fig60Hours}a, across a 3 day period the temperature varied by $\Delta T$$\approx$0.4~K.  The voltage exhibited a peak-to-peak fluctuation of $\Delta V$$\approx$6~ppm.  No correlation between voltage and temperature could be determined.

\par A histogram of measured voltage, Fig.~\ref{fig60Hours}b, shows a fairly Gaussian distribution indicative of true noise.  The full-width at half-maximum variation was 2.28(4)~ppm over the entire 3~day measurement, while the standard deviation across the entire 3~day period was $\sigma(V)$=1.22~ppm.  Based on our full-scale MRTOF-MS \cite{Schury2014}, the mass resolving power achievable using such voltage stability would be limited to $R_\textrm{m}<$7.4$\times$10$^5$.  This is compatible to even the most stringently demanding isobar separation applications.


 
\par For use as a mass analyzer, the application of drift correction methods makes the intermediate term ($\sim$10~minutes) stability the limiting factor.  Figure~\ref{fig60Hours}c shows the measured voltage variance histogram for all consecutive 10~minute intervals within the 3~day measurement.  The average full-width at half-maximum variation across all consecutive 10~minute intervals was 2.08(51)~ppm.  Thus, the long-term and intermediate-term performance are not very dissimilar.  With this level of intermediate stability, the mass resolving power achieved by an infinitely-high-resolution device would be limited to $R_\textrm{m}<$8.4$\times$10$^5$, based on the performance of our full-scale MRTOF-MS.

\subsection{Intermediate-term stability as a function of gain and output voltage}
\par
\begin{figure}[b]
	\vspace{3mm}
	\centering
	 \includegraphics[width = 3.45in]{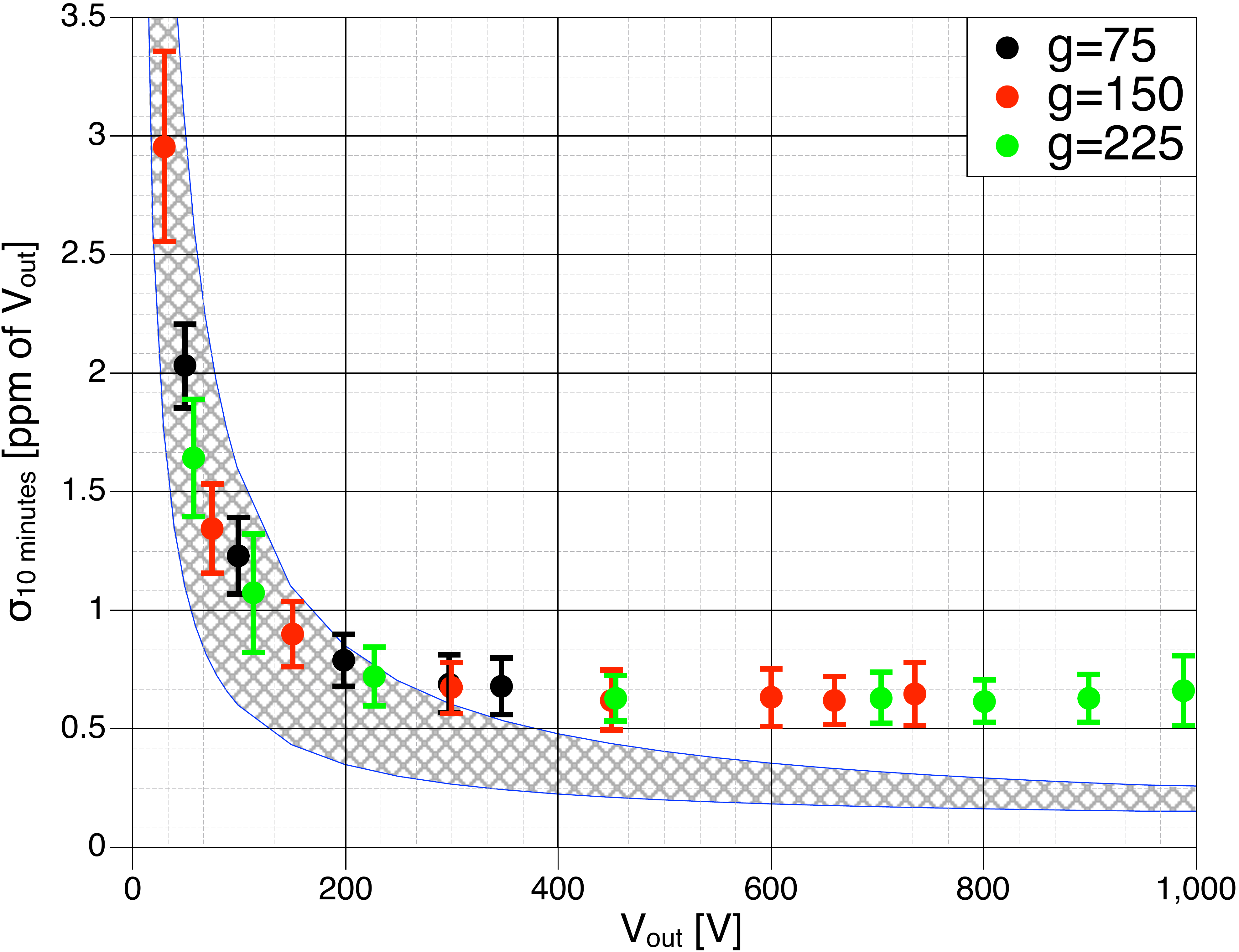}
	\caption{\label{figStab2} Intermediate-term stability as functions of output voltage for multiple total gain values.  Each point is the result of at least 3 hours digitization at 1.7~Hz.  The hashed region represents the noise floor of the measurement device.
	} 
\end{figure}

 In order to further evaluate the performance of the device, stability measurements were made using three gain values at output voltages up to 1~kV.  The tests were limited to 1~kV by the maximum allowable voltage of the Agilent 3458A digital multimeter.  For each point in Fig.~\ref{figStab2}, the output value was set and the device was given several minutes to settle and achieve thermal equilibrium.  The voltage was then digitized for at least 3 hours at 1.7~Hz as done previously.  The data were separated into 10 minute sequences and the standard deviation of each sequence was calculated.  The value and uncertainty are given by the average and variance, respectively, of standard deviations across all 10 minute sequences.

The noise floor of the 3458A multimeter operated at 1.7~Hz in the 1~kV measurement range, is represented by the hashed area in Fig.~\ref{figStab2}.  When using 600~ms integration time, the multimeter's datasheet indicates a noise floor that is $\approx$0.1~ppm of the voltage reading with an additional $\approx$50~$\mu$V$_\textrm{RMS}$ noise level on the 1~kV measurement range, which is responsible for the noise floor following a reciprocal curve shape in Fig.~\ref{figStab2}.   The datasheet further suggests a threefold span between RMS noise and ``peak RMS noise" that is reflected in the span of the hashed region.

 
\subsection{Short-term stability and noise}
\par While a digitization rate of 1.7~Hz was used to achieve the requisite precision needed for the study of long- and intermediate-term stability, this digitization frequency precluded investigation of the contribution from 50~Hz mains noise.  As our MRTOF-MS is typically operated with a 15~ms cycle, fluctuations at frequencies up to $\approx$100~Hz may cause peak broadening.  To study the device performance in such frequency ranges, the 3458A was set to a 2~ms integration period to allow the voltage to be digitized at 500~Hz.  At this rate, however, we were unable to reliably trigger the digitization using the computer code running on the Raspberry Pi and thus unable to continuously stream the digitized data.  Rather, we used the 3458A's internal triggering option to trigger a digitization every 2~ms.  For technical reasons, only 5000 digitizations at full precision could be performed in a single burst, allowing for 10~s digitization periods, with $\approx$500~ms between each burst.  The digitized data were transferred to the control computer during the time between each burst.  

\begin{figure}[b]
	\vspace{-0.3mm}
	\centering
	 \includegraphics[width = 3.47in]{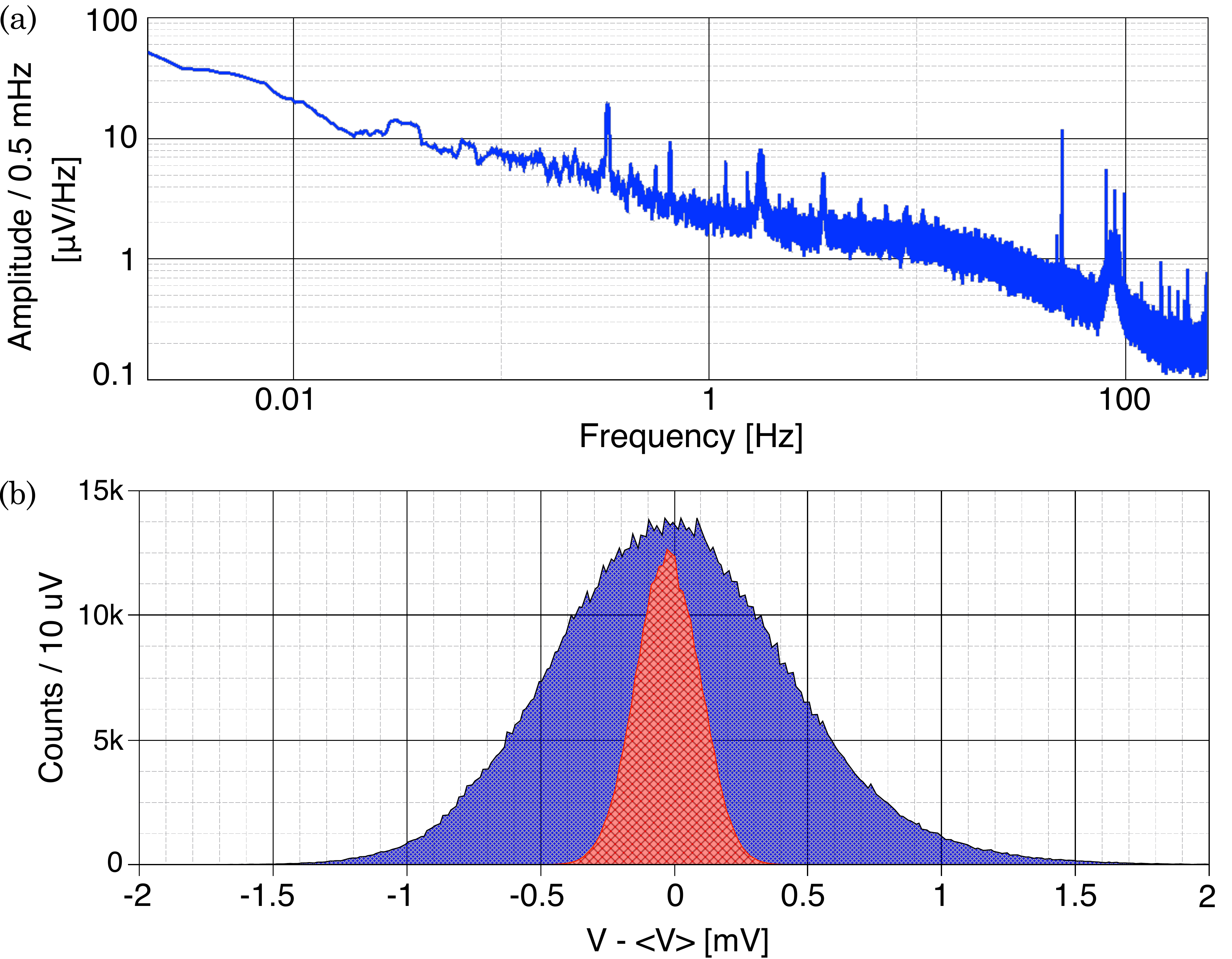}
	\vspace{-5mm}
	\caption{\label{figFFT} (a) Fast Fourier transform (FFT) of a 100~V signal output from a $g$=999 voltage stabilization circuit.  (b) Histogram of the 10$^6$ data used in the FFT is shown in purple.  In red is a histogram from a measurement made under similar conditions, but with the input leads of the 3458A shorted together.  The 100~V signal has a FWHM of 950~$\mu$V, while the inherent noise level is seen to have a FWHM of 250~$\mu$V. 
	} 
\end{figure}

\par We collected data from consecutive 10~s digitization bursts until 10$^6$ data points were accumulated, requiring approximately 35~minutes.  This was done for output voltages of 0~V, 10~V, and 100~V with power supplies of gains $g$=100 and $g$=999.  Fast Fourier transforms (FFT) were made from each data set.  The results from a $g$=999 circuit with 100~V output is shown in Fig.~\ref{figFFT}.   

\par The amount of 50~Hz signal noise as well as the width of the voltage measurement histogram were independent of the power supply gain or output voltage.  We draw two conclusions from this invariance.  First, the majority of 50~Hz signal noise is likely introduced after the voltage regulation circuitry, as we would expect to see a 10-fold increase in the signal amplitude when comparing gains $g$=100 and $g$=999 were the noise propagated through the amplification stages.  Second, the observed Gaussian distribution is most likely the result of the digitization precision limit of the 3458A with 2~ms integration time, as the width would otherwise be expected to scale with the output voltage.  However, it cannot be completely ruled out that there may be $\approx$1~mV$_\textrm{rms}$ of wideband noise on the output independent of the voltage or gain.

\section{Initial testing with MRTOF-MS}
\par As our presently operational full-scale MRTOF-MS\cite{Schury2014} could not be used for this study due to a need for maintaining online operational readiness, we have tested the power supplies on a recently constructed half-scale device.  This MRTOF-MS, as with the full-scale version, consists of two electrostatic mirrors.  For the injection side mirror we use three electrodes to create a hard mirror, while for the ejection side mirror seven electrodes are used to create a soft mirror.  A lens electrode is located just before the innermost ejection-side mirror electrode.  The total length, from the outermost edge of the first injection-side electrode to the outermost edge of the final ejection electrode is $\approx$45~cm.  A highly detailed description of the configuration and operation of the full-scale version can be found in a previous manuscript\cite{Schury2014}.

\par A crate of 16 power supplies was built to bias the MRTOF-MS electrodes.  The voltages were adjusted until a time resolving power of $\frac{t}{\Delta t}\approx$250,000 was achieved.  The voltages applied to each electrode and the relative effect of each electrode on the time-of-flight are shown in Table~\ref{tabBiases}.  Due to the lens voltage requiring a bias of 5~kV or more, we have thus far opted to use an NHR~42~60r high-precision $\pm$6 kV power supply from ISEG to bias this electrode.

\begin{table}[htp]
\caption{Voltages applied to each electrode and the relative effect a relative variation in the voltage has on the time-of-flight.  The ejection-side mirror electrodes are are designated ``Ej" while the injection-side mirror electrodes are designated ``Inj".  Electrode number increases toward the drift tube.}
\begin{center}
\begin{tabular}{c | c | c}
Name & Voltage [V] & $\frac{\Delta t}{t}/\frac{\Delta V}{V}$ [ppm/ppm] \\
\hline
Inj0 (open) & -920 & 1.1$\times$10$^{-4}$ \\
Inj0 (closed) & 377.197 & -0.055 \\
Inj1 & -173.001 & 0.003 \\
Inj2 & 133.449 & 0.001 \\
Drift tube & -1505.4 & -0.292 \\
Lens$^\textrm{a}$ & -5001 & -0.051 \\
Ej6 & -848.663 & -0.016 \\
Ej5 & -739.5108 & -0.015 \\
Ej4 & -454.612 & -0.013 \\
Ej3 & -633.109 & -0.015 \\
Ej2 & 114.898 & -0.002 \\
Ej1 & -181.301 & 0.008 \\
Ej0 (open) & -202 & -3$\times$10$^{-5}$ \\
Ej0 (closed) & 699.996 & -0.057 \\
\hline
\end{tabular}
\footnotetext{ISEG NHR 42 60r High-Precision $\pm$6 kV Power Supply}
\end{center}
\label{tabBiases}
\end{table}%

\par The MRTOF-MS was tested with $^{133}$Cs$^+$ making $n$=290~laps.  The system was operated with a 15~ms cycle.  As the spectral peaks had a median Gaussian width of $\sigma$$\approx$8~ns, the ion source was adjusted to deliver approximately one detected ion per cycle, on average, to preclude uncounted ions from detector deadtime.  With the voltage distribution of Table~\ref{tabBiases}, the voltage applied to each electrode was varied in a $\pm$1~V range to verify a locally linear response and determine the relative effect that a relative variation in the voltage has on the time-of-flight.  

\begin{figure}[t]
	\vspace{3mm}
	\centering
	 \includegraphics[width = 3.45in]{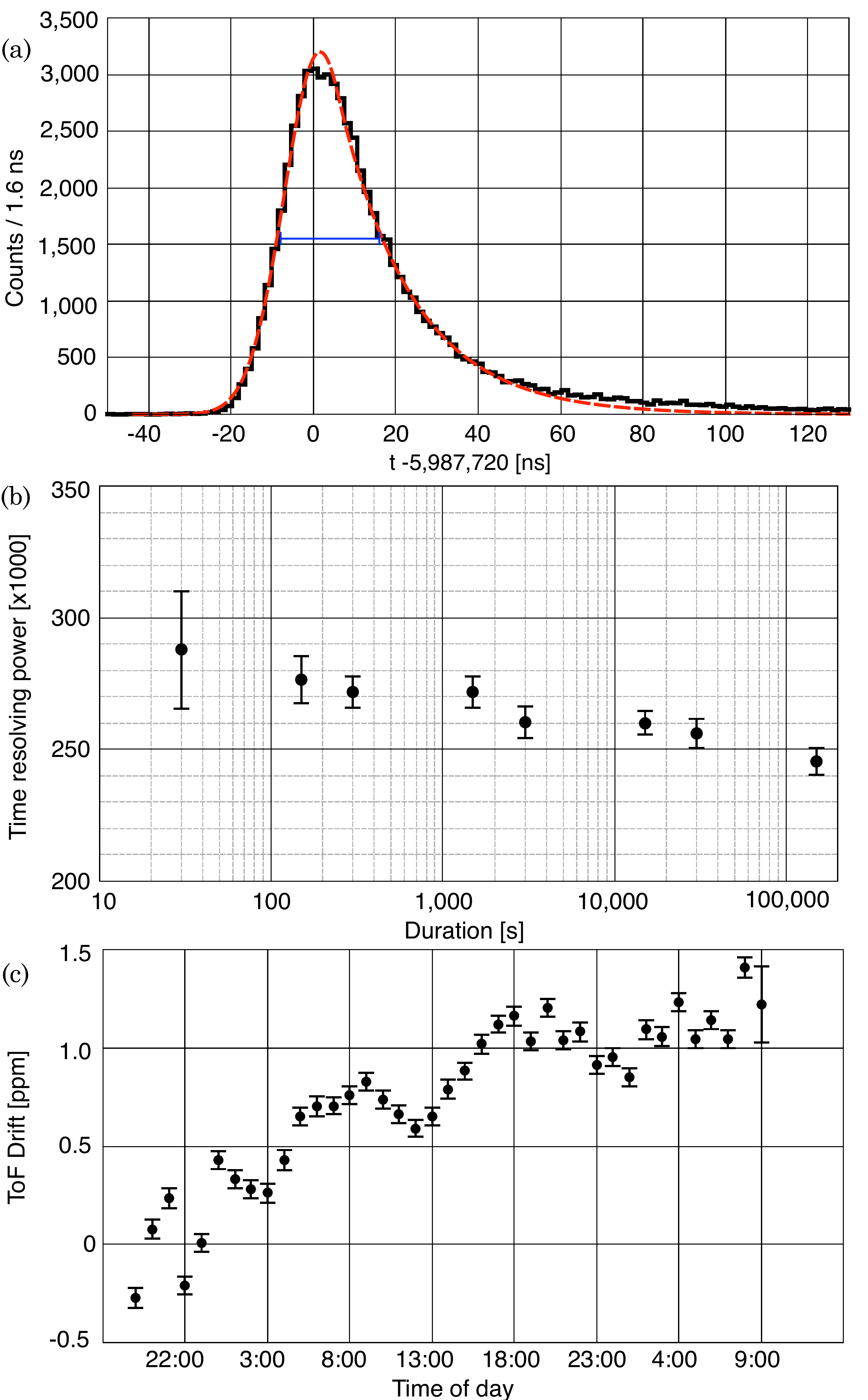}
	\caption{\label{figMRTOF_all} Results of long-term stability test of MRTOF-MS.  The MRTOF-MS was operated at a time resolving power of $\frac{t}{\Delta t}\approx$250,000 for 36 hours.  Panel (a) shows the spectral peak data and the fit curve from the first hour of data accumulation.  The blue line indicates the FWHM from which the time resolving power is determined.  Panel (b) shows the time resolving power as a function of data accumulation duration.  Panel (c) shows the relative drift in the time-of-flight peak center over the 36 hours.  The first hourlong accumulation period served as the reference time-of-flight.  Each data point in panel (c) is the difference between the peak center from one consecutive hourlong set of data and the reference time-of-flight.
	} 
\end{figure}

\par Similar to the benchtop test of the power supplies, the ion time-of-flight was measured for 36 hours.  The results are shown in Fig.~\ref{figMRTOF_all}.  The spectral peak exhibits a not insignificant slow tail which is typical.  Better tuning of the mirror potentials should eventually suppress this feature.  Figure~\ref{figMRTOF_all}(b) shows the time resolving power as a function of the measurement duration.  This was determined by fitting the data from first cycle through to the listed time; the final point is the complete 36 hour data set.  The reason that the time resolving power decreases with longer measurement duration can be gleaned from Fig.~\ref{figMRTOF_all}(c), which shows the time-of-flight drift in consecutive hourlong measurements.  The system exhibits a drift of less than 2~ppm in 36 hours, but this is enough to reduce the time resolving power by 10\% over the course of 36~hours.  The observed drift indicates that the MRTOF drift tube potential could have experienced no more than 7~ppm voltage drift over the 36 hours of the measurement.  It should be noted here that in addition to possible instabilities in the power supplies, the MRTOF-MS itself is susceptible to thermal expansion which is expected to produce time-of-flight shifts of $\sim$10~ppm/K.

\par Our full-sized MRTOF-MS has been operated with voltages stabilized by a digital PID control loop\cite{WadaPID} from its inception.  We recently added RC filters (2.5~M$\Omega$/20~$\mu$F) to reduce the voltage fluctuations as suggested by the results of Wienholtz\cite{Wienholtz2019}.  In a similar way, we also used the Allan deviation\cite{R_AllanVar} to make a quantified comparison in the performance of the original PID-based stabilization, the RC filter enhanced PID-based stabilization, and the stability of the circuit presented here.  Rather than perform consecutive peak fittings, however, we have calculated the Allan deviation of the individual detected ions.  In each detection cycle, the average time-of-flight of all detected ions is calculated; cycles without detected ions are omitted.  The Allan deviation is then calculated based on this sequence of data, with the minimum Allan deviation period being on the order of the cycle period -- due to empty cycles, the actual minimum is greater, however. 
\par
\begin{figure}[t]
	\vspace{3mm}
	\centering
	 \includegraphics[width = 3.45in]{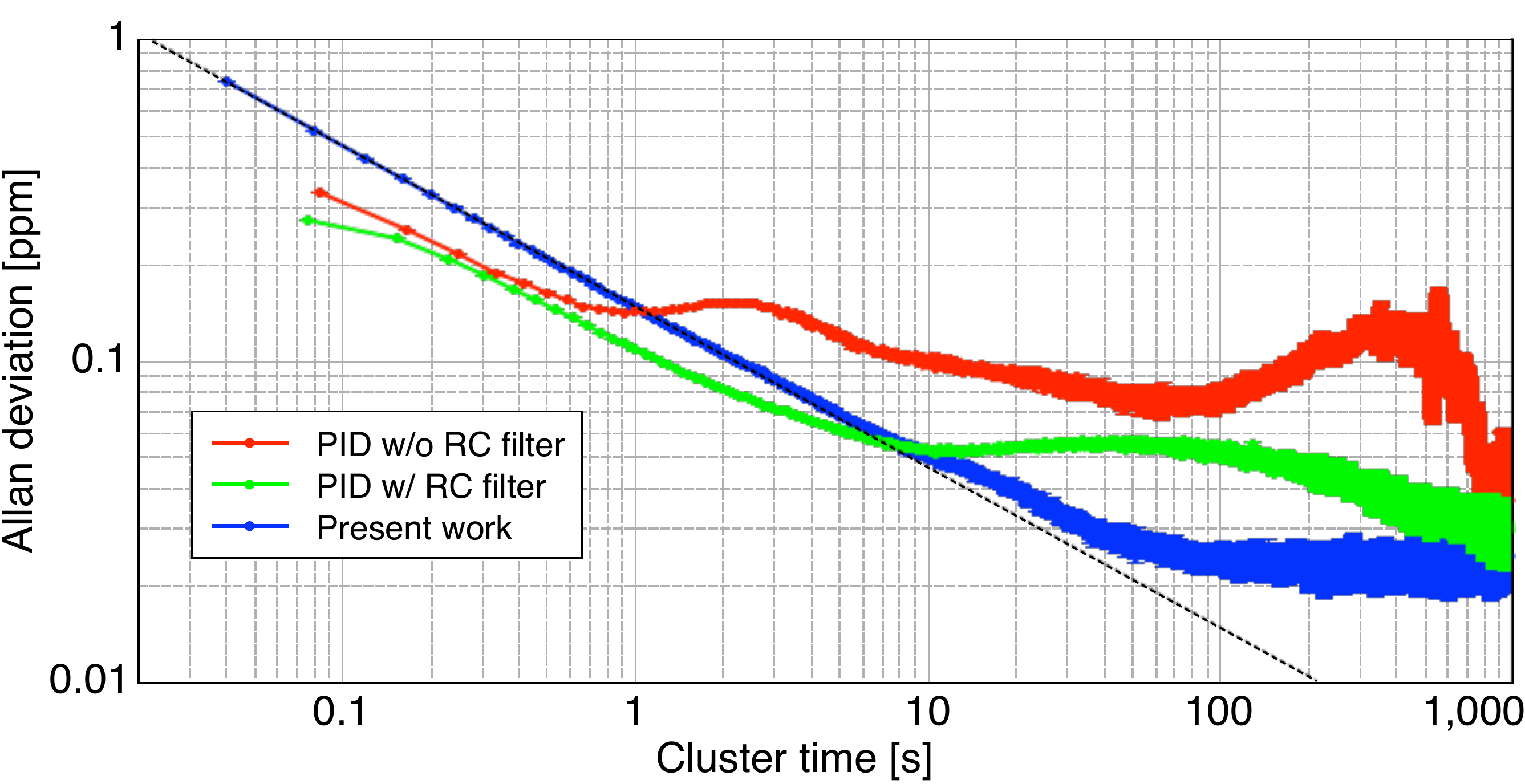}
	\caption{\label{figAllanMRTOF} Allan deviation from individual ion times-of-flight.  The PID regulation without and with RC filter exhibit a slope approximating white noise to $\approx$1~s and $\approx$10~s, respectively.  The circuitry of the present work exhibits a slope consistent with white noise\cite{IEEE2008} as shown by the dashed line.  The Allan deviation curve switches to a flat line indicative of flicker noise at a level of $\approx$0.02~ppm, which coincides with the 0.1~ns raw bin width from the TDC.  The absolute values of the Allan deviation are initially lower for the PID-based data because the full-size MRTOF has been tuned to a time resolving power of $\sim$6$\times$10$^{5}$, while the newer half-sized device has only been tuned for a time resolving power of $\sim$2.5$\times$10$^{5}$.
	} 
\end{figure}

 Figure~\ref{figAllanMRTOF} shows the result of this analysis for the data set used in Fig.~\ref{figMRTOF_all} along with a similar analysis using pre-existing shorter duration data from our full-size MRTOF-MS.  The data for the PID stabilization was obtained using a 30~ms cycle period, yielding a minimum Allan deviation sampling period of $\approx$80~ms.  The data for the present system was obtained using a 15~ms cycle period, yielding a minimum Allan deviation sampling period of $\approx$40~ms.  This analyses shows that in terms of long-term stability the voltage stabilization circuit presented exceeds the performance of our RC filter enhanced PID regulation system.  Using the power supplies presented in this work, the MRTOF-MS exhibits white noise in the ion time-of-flight over a much wider duration than with the PID-based regulation systems.

\section{Conclusion and outlook}
We have designed and bench-tested a high-voltage regulation circuit which settles on the ppm-level within one minute and has proven capable of maintaining $\sigma$($V$)/$V$$\approx$1~ppm for several days.  The device uses high-voltage photodiodes and high-flux infrared LEDs in combination with a high-voltage 1000:1 resistive voltage divider and low-noise op-amp IC to produce a high-voltage, $g$=$-$999 op-amp.  A secondary gain-adjust amplifier produces a net-positive gain which can be adjusted without the need to replace the expensive high-voltage divider.  A 20-bit DAC with extremely low-noise voltage reference drives the op-amp.  The high-voltage output was measured to have an integral linearity comparable to the 20-bit DAC itself.
\par Over the course of 3~days of evaluation at 400~V output, the maximum relative deviation was observed to be $\approx$6~ppm, while the standard deviation was 1.22~ppm across the evaluation period.  Thermal effects were determined to be very low, with a temperature dependent offset voltage of -307(2)~$\mu$V/K and an inferred gain temperature coefficient of $-$0.6~ppm/K.  While limitations in test equipment precluded characterization of outputs exceeding 1~kV, the flat response in the stability as a function of output voltage gives us some confidence that the device will provide $\sim$1~ppm stability up to the design limit of $\pm$5~kV.
\par High-speed digitization at a reduced precision confirms that the device has a very low amplitude noise spectrum in the 0.1~Hz--100~Hz range, with the measured 50~Hz noise likely being from external environmental sources.  This is further confirmed with calculations of the Allan deviation for ion time-of-flight measurements performed at a 66~Hz repetition rate.  Such calculations also indicate that this voltage stabilization circuit provides a long-term stability exceeding that of a long time constant RC filter enhanced PID stabilization circuit.
\par We have applied these power supplies to a newly constructed MRTOF-MS which is a half-scale copy of our previously existing system.  We could maintain a time resolving power of 250,000 over 36 hours.  The maximum time-of-flight peak drift over 36 hours was less than 2~ppm.  This is sufficient for even the more demanding of mass separator performance.  The measured performance of the power supplies across all time domains indicate that they should allow for eventually achieving a mass resolving power approaching 10$^6$, although such resolving power would still require the use of ToF drift correction techniques\cite{Schury2018, Fischer2018}.


This work has been supported by the Japan Society for the Promotion of Science KAKENHI (Grant Numbers 2200823, 24224008, 24740142, and 15K05116).   We wish to express gratitude to R. Molzon of Michigan Technological University for assistance in the statistical analyses.

\end{document}